\documentclass[journal]{IEEEtran}

\ifCLASSINFOpdf
\else
   \usepackage[dvips]{graphicx}
\fi
\usepackage{url}

\usepackage{graphicx}
\usepackage{amssymb}
\usepackage{mathrsfs}
\usepackage{amsmath}
\usepackage{cite}
\usepackage{caption}
\usepackage{bm}
\usepackage{booktabs}
\usepackage{subfigure}
\usepackage{float}
\usepackage{multirow}
\usepackage{xcolor}

\begin{document}
\captionsetup{font={small}}

\title{A Refining Underlying Information Framework for Monaural Speech Enhancement}

\author{Rui Cao, Tianrui Wang, Meng Ge$^{*}$, Longbiao Wang$^{*}$, \IEEEmembership{Member, IEEE}, and Jianwu Dang, \IEEEmembership{Member, IEEE}
\thanks{Rui Cao and Tianrui Wang are with the Tianjin Key Laboratory of Cognitive Computing and Application, College of Intelligence and Computing, Tianjin University, Tianjin 300354, China (e-mail: caorui\_2022@tju.edu.cn; wangtianrui@tju.edu.cn).}
% liumeng2017@tju.edu.cn).}
\thanks{Meng Ge is with the Saw Swee Hock School of Public Health, National University of Singapore, Singapore 117549 (e-mail: gemeng@nus.edu.sg).}
% \thanks{Weibin Zhu is with the Institute of Information Science and the Beijing Key Laboratory of Advanced Information Science and Network Technology, Beijing Jiaotong University, Beijing 100044, China (e-mail: wbzhu@bjtu.edu.cn).}
\thanks{Longbiao Wang is with the Tianjin Key Laboratory of Cognitive Computing and Application, College of Intelligence and Computing, Tianjin University, Tianjin 300354, China, and also with Huiyan Technology (Tianjin) Company, Ltd., Tianjin 300384, China (e-mail: longbiao\_wang@tju.edu.cn).}
\thanks{Jianwu Dang is with the Tianjin Key Laboratory of Cognitive Computing and Application, College of Intelligence and Computing, Tianjin University, Tianjin 300354, China, and also with the Pengcheng Laboratory, Shenzhen 518000, China (e-mail: jdang@jaist.ac.jp). $^{*}$ Corresponding author.}
}

\maketitle
\begin{abstract}
Supervised speech enhancement has gained significantly from recent advancements in neural networks, especially due to their ability to non-linearly fit the diverse representations of target speech, such as waveform or spectrum. However, these direct-fitting solutions continue to face challenges with degraded speech and residual noise in hearing evaluations. By bridging the speech enhancement and the Information Bottleneck principle in this letter, we rethink a universal plug-and-play strategy and propose a Refining Underlying Information framework called RUI to rise to the challenges both in theory and practice. Specifically, we first transform the objective of speech enhancement into an incremental convergence problem of mutual information between comprehensive speech characteristics and individual speech characteristics, e.g., spectral and acoustic characteristics. By doing so, compared with the existing direct-fitting solutions, the underlying information stems from the conditional entropy of acoustic characteristic given spectral characteristics. Therefore, we design a dual-path multiple refinement iterator based on the chain rule of entropy to refine this underlying information for further approximating target speech. Experimental results on DNS-Challenge dataset show that our solution consistently improves 0.3+ PESQ score over baselines, with only additional 1.18 M parameters. The source code is available at \url{https://github.com/caoruitju/RUI_SE}.
\end{abstract}
\IEEEpeerreviewmaketitle
\section{Introduction}
\IEEEPARstart{M}{onaural} speech enhancement focuses on extracting the target speech from the corresponding noisy recording, aiming to improve both the quality and intelligibility of speech. Traditional speech enhancement methods usually rely on mathematical formulas derived from assumptions about the statistical characteristics of speech and noise, such as spectral subtraction \cite{paliwal2010single}, Wiener filtering \cite{chen2006new} and subspace-method \cite{asano2000speech}. Recent data-driven speech enhancement methods have achieved significant performance gains by employing neural networks to fit the non-linear mapping relationship between the input noisy speech and the target speech, avoiding the statistical assumptions in traditional methods. The existing data-driven methods can be categorized into two streams, namely the time domain and the (time-frequency)-domain. Time-domain methods \cite{luo2019conv, guimaraes2020monaural, kong2022speech} aims to model the distribution of waveform samples via convolution-based speech encoding and decoding operations, while (time-frequency)-domain methods \cite{loizou2005speech, wang2014training,paliwal2011importance,tan2019complex, tan2019learning, hu20g_interspeech} work by separating noisy complex spectrum into either magnitude-phase or real-imaginary components to minimize the distance to the complex spectrum of target speech. However, these data-driven direct-fitting methods continue to face difficulties in addressing speech over-suppression and noise under-suppression \cite{jiang2022speech,kong23c_interspeech}, which can negatively impact hearing evaluations, especially in challenging acoustic scenarios.

Recent related studies \cite{yin2020phasen, wang2022hgcn, wang2022harmonic, he2023speech} have shown that preserving acoustic characteristics inherent in speech (e.g., articulatory attributes, acoustic structural features, and auditory perceptual attributes\cite{coker1976model, browman1992articulatory,welling1998formant,moore1983suggested,kent2018static,plourde2008auditory}) can ensure that enhanced speech remains consistent with human hearing perception. These acoustic characteristics are closely linked to speech production and reflect the fundamental characteristic of natural human speech. Correcting these acoustic structures in speech processing makes the enhanced speech sound more natural and familiar to human listeners. Motivated by this fact, we rethink the speech enhancement task and prompt a research question: Could we theoretically repair the incomplete intrinsic characteristics in enhanced speech (e.g., articulatory attributes) to perfectly approximate the hearing perception on target speech?

To answer this research question, we propose a universal hearing-repair speech enhancement framework, called \textbf{R}efining \textbf{U}nderlying \textbf{I}nformation (RUI). Our RUI is inspired by the Information Bottleneck principle \cite{tishby2015deep}. We find similarities between speech enhancement task and Information Bottleneck principle, with both primarily aiming to compress non-target information (i.e., noises) while capturing the most relevant information (i.e., target speech) for the target object. Motivated by this, we transform the objective of speech enhancement into an incremental convergence process of mutual information among speech characteristics shown in Fig. \ref{venn}, including spectral characteristic $\mathbb{P}$, acoustic characteristic $\mathbb{A}$, and comprehensive characteristic $\mathbb{C}$. Thus, it can be further derived that the optimization objective is essentially the sum of the entropy $H(\mathbb{P})$ and the conditional entropy $H(\mathbb{A|P})$. Existing direct-fitting methods pay more attention to how to model $H(\mathbb{P})$, while our solution further explore the underlying information originating from $H(\mathbb{A|P})$. This theoretically answers how our proposed RUI can achieve the repair of characteristics inherent in target speech. 

Based on the above discussion, we employ a pre-enhancement module and an underlying information extractor to explore $H(\mathbb{P})$ and $H(\mathbb{A})$, respectively. Additionally, we iteratively expand the conditional entropy according to the chain rule of entropy and design a multiple refinement iterator through dual-path residual mechanism for modeling $H(\mathbb{A|P})$. By doing so, the intrinsic characteristics in target speech are asymptotically refined in the output-enhanced speech, gradually improving the hearing perception in practice.

% This letter is organized as follows. In Section \ref{sec:method}, we motivate and design the proposed RUI via Information Bottleneck principle. In Section \ref{sec:experiment}, we report the experiments. Section \ref{sec:conclusion} concludes the study.
% \vspace{-1em}
\section{Methodology}
\label{sec:method}
\subsection{Information Bottleneck in Speech Enhancement}
The Information Bottleneck principle suggests that DNN should learn to extract the most efficient informative representation in the input variable about the output-label-variable and maximally compress irrelevant representation. Overall, it exhibits significant consistency with speech enhancement, where the goal is to extract the target speech from the noisy speech and suppress noise as much as possible \cite{das2021fundamentals}. Derived from the Information Bottleneck principle, the universal optimization process of DNN-based speech enhancement can be formulated as the minimization of the following Lagrangian,
\begin{equation}
    \setlength{\abovedisplayskip}{4pt}
    \setlength{\belowdisplayskip}{4pt}
    \label{1}
        \mathcal{L} = I({X;\hat{S}}) - \beta I({\hat{S};S})
\end{equation}
where $X$, $\hat{S}$, and $S$ represent noisy speech, enhanced speech, and clean speech, respectively. $I(\cdot ; \cdot)$ denotes mutual information. $\beta$ represents the positive Lagrange multiplier. Given the restrictions of accurately modeling the infinite noise, minimizing the gap between $\hat{S}$ and $S$ is typically formulated as the optimization objective $\Gamma$ of supervised speech enhancement \cite{saleem2019review}. However, as a bond between the suppression of noise $I({X;\hat{S}})$ and the recovery of clean speech $I({\hat{S};S})$, $\hat{S}$ is subjected to the unclear boundary between speech and noise during the trade-off process, causing degraded speech and residual noise. We hold that the occurrence of these issues is due to a lack of effective utilization of acoustic characteristics of speech, as speech possesses distinct acoustic characteristics that differentiate it from noise \cite{wang2022hgcn}. Explicitly modeling them contributes to facilitating the formation of speech-specific feature boundary. Therefore, we delve deeply into a universal way of incorporating the acoustic characteristic (e.g., articulatory attributes) inherent in speech signals. Essentially, we extend the recovery of clean speech to the incremental convergence of the mutual information between the characteristic information $\bm{c}$ of speech, 
\begin{equation}
    \setlength{\abovedisplayskip}{4pt}
    \setlength{\belowdisplayskip}{4pt}
    \label{2}
          \mathcal{F}(X;\mathbb{P})=\bm{c}^{\hat{S}},\mathcal{F}(X;\mathbb{C})=\bm{c}^{S}
    \vspace{-0.06cm}
\end{equation}
\begin{equation}
    \setlength{\abovedisplayskip}{4pt}
    \setlength{\belowdisplayskip}{4pt}
    \label{3}
         maximize \quad \Gamma = I({\bm{c}^{\hat{S}};\bm{c}^S})
    \vspace{-0.06cm}
 \end{equation}
where $\mathcal{F}(\cdot;\mathbb{P})$ denotes the regression DNN $\mathcal{F}$ that performs the optimization process in Eq.~(\ref{1}) only in conjunction with spectral characteristic $\mathbb{P}$. However, the existing spectrum estimation approaches are inadequate to achieve the comprehensive characteristic $\mathbb{C}$ of clean speech. This incompleteness of characteristic results in the optimization objective converging to a supremum,
\begin{equation}
    \setlength{\abovedisplayskip}{4pt}
    \setlength{\belowdisplayskip}{4pt}
    \label{4}
    sup\ \Gamma=I(\mathbb{P};\mathbb{C})
    \vspace{-0.06cm}
\end{equation}

\begin{figure}[t]
\vspace{-1.5em}  
\centerline{\includegraphics[width=5cm]{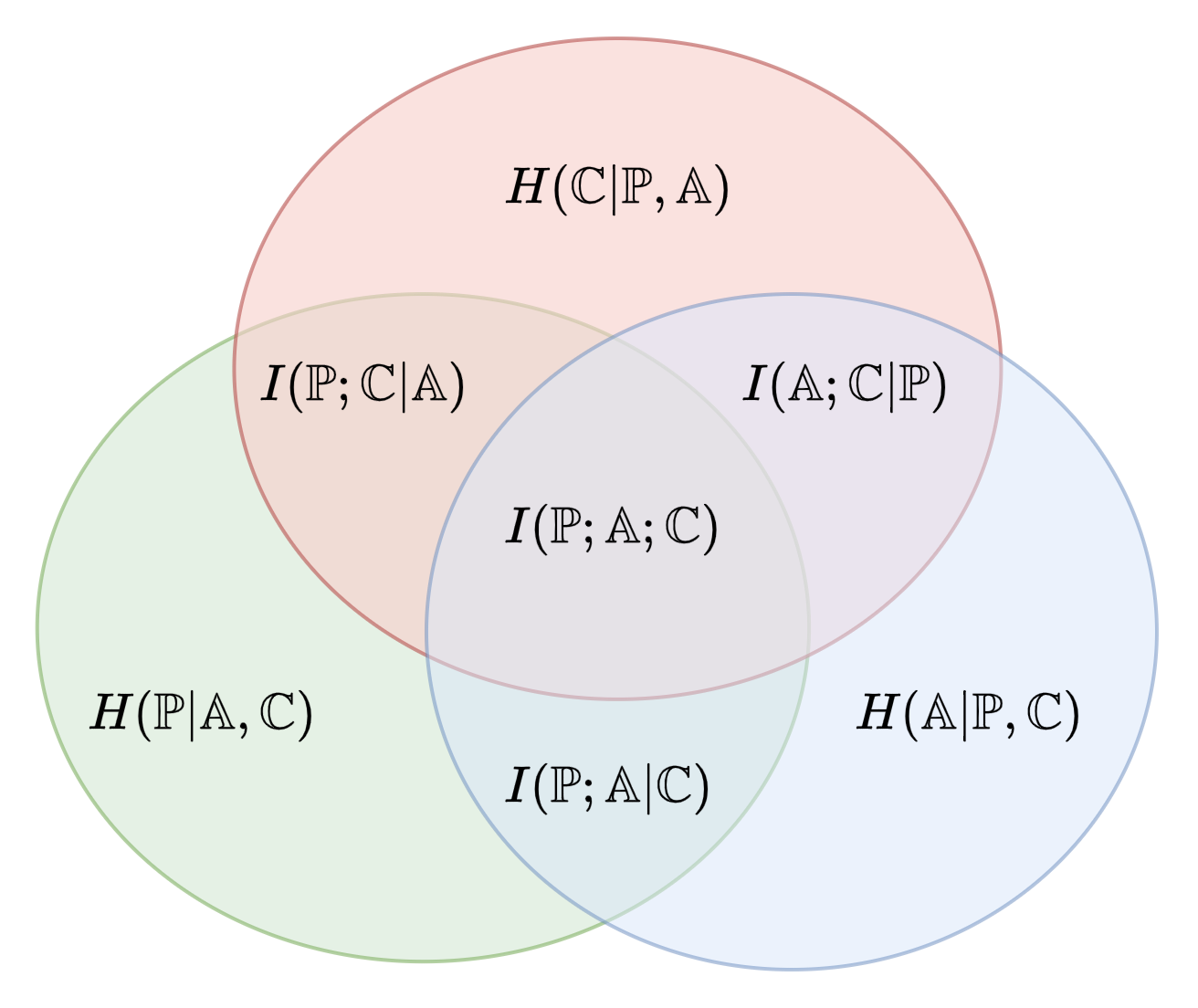}}
\captionsetup{justification=justified}
\vspace{-0.6em}  
\caption{Venn diagram illustrating specific relationships between each characteristic. The region overlapping with the red circle represents the upgraded upper bound $\Gamma_{u}$. The green, blue, and red circles represent characteristics $\mathbb{P}$, $\mathbb{A}$, and $\mathbb{C}$.}
\label{venn}
\vspace{-1.8em} 
\end{figure}

The acoustic characteristic inherent in speech signals is represented as $\mathbb{A}$. The specific interrelationships between each characteristic are illustrated in Fig.~\ref{venn}. From the perspective of incremental convergence, the upgraded upper bound $\Gamma_{u}$ after explicitly incorporating $\mathbb{A}$ can be formulated as,
\begin{equation}
    \setlength{\abovedisplayskip}{4pt}
    \setlength{\belowdisplayskip}{4pt}
    \begin{split}
        \label{5}
        \Gamma_{u}
        & = I(\mathbb{P};\mathbb{C}|\mathbb{A}) + I(\mathbb{P};\mathbb{A};\mathbb{C}) +
        I(\mathbb{A};\mathbb{C}|\mathbb{P}) \\
        & = I(\mathbb{P};\mathbb{C}) + I(\mathbb{A};\mathbb{C}|\mathbb{P})
    \end{split}
\end{equation}

We formally define the conditional mutual information $I(\mathbb{A};\mathbb{C}|\mathbb{P})$ as underlying information in this letter. This is a crucial element for approximating the optimal information theoretic limit of the optimization. The direct introduction of underlying information in $\hat{S}$ also goes a step further to promoting Eq.~(\ref{1}), for enhancing the essence of speech and eliminating noise impurity. 

\begin{figure*}[htb]
    \centering
    \vspace{-0.1cm}
    \includegraphics[width=16cm]{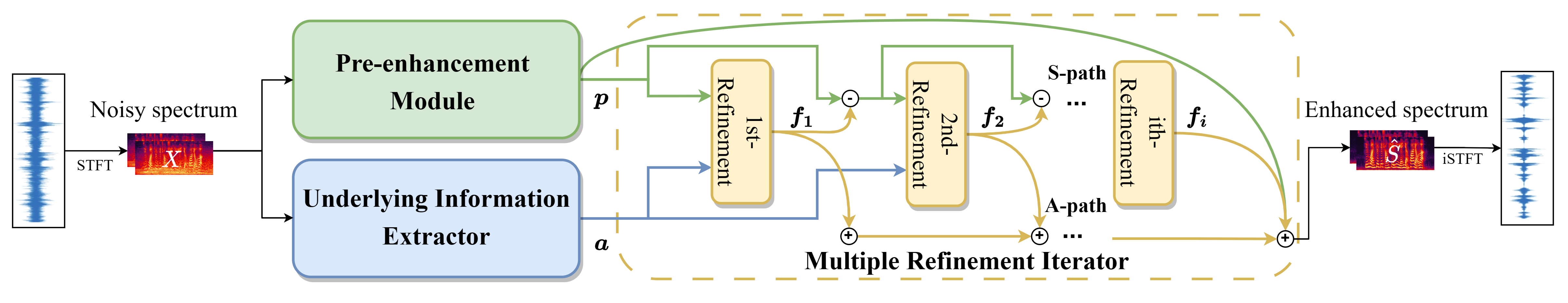}
    \captionsetup{justification=centering}
    \vspace{-0.4em}  
    \caption{The proposed Refining Underlying Information Framework.}
    \label{ICRF}
    \vspace{-0.65cm}
\end{figure*}
\subsection{Refining Underlying Information Framework}
For the feedforward computation, the mutual information in Eq.~(\ref{5}) degenerates into entropy. Decoupling the spectrum has been validated as an effective strategy for obtaining sparse term \cite{li2021icassp, li2022glance,ijcai2022p582,li2022filtering}. Motivated by this, to thoroughly refine the underlying information, we iteratively expand the second term according to the chain rule of entropy, which can be formulated as,
\begin{equation}
    \setlength{\abovedisplayskip}{4pt}
    \setlength{\belowdisplayskip}{4pt}
\label{6}
    \mathcal{F}_{u}(X) = H(\mathbb{P}) + H(\mathbb{A|P}) = H(\mathbb{P}) + \sum^{N}_{i=1} H(\mathbb{A}_{i}|\mathbb{P} ,..., \mathbb{A}_{i-1})
\end{equation}

To indicate the direction of information in the framework, we use the concept of information flow \cite{goldfeld2019estimating} as an intuitive alternative of Shannon entropy. In this way, we design a refining underlying information framework (RUI) as shown in Fig.~\ref{ICRF}, and the forward flow $\mathcal{F}_{u}(X)$ can be further parameterized as follows,
\begin{equation}
    \setlength{\abovedisplayskip}{4pt}
    \setlength{\belowdisplayskip}{4pt}
\label{7}
    \mathcal{F}_{u}(X) = \bm{p} + \sum^{N}_{i=1} R_{i}(\bm{a},\bm{p}-\sum^{i-1}_{j=1}\bm{f}_{j})
\end{equation} 
where $\bm{p}$, $\bm{a}$, and $\bm{f}_i$ represent the information flows after passing through the pre-enhancement module (PEM), underlying information extractor (UIE), and $i$-th refinement $R_{i}$ of multiple refinement iterator (MRI). $\bm{p}-\sum^{i-1}_{j=1}\bm{f}_{j}$ imitates the condition term of conditional entropy in Eq.~(\ref{6}). 

The input of RUI is the noisy complex spectrum computed by Short-Time Fourier Transform (STFT), denoted as $X\in \mathbb{R}^{T\times 2 \cdot F}$, where $T$ and $2 \cdot F$ denote the number of time frames and frequency bins (real and imaginary part). Any complex-spectrum approach can serve as the PEM, and its result $\bm{p} \in \mathbb{R}^{T\times 2 \cdot F}$ is considered as a preliminary enhanced output. To explicitly repair the incomplete articulatory attributes in PEM, our previously proposed Harmonic Attention\footnote{\url{https://github.com/caoruitju/RUI_SE/blob/main/HA/HA.md}} \cite{wang2023harmonic} is employed as the UIE to actively capture the comb-like harmonics based on the correlation between the noisy spectrum and a comb-pitch conversation matrix, resulting in the flow $\bm{a}\in \mathbb{R}^{T\times C \times F}$ with the highlighted acoustic structural information. Finally, MRI performs structured correction of flow $\bm{p}$ with the assistance of flow $\bm{a}$, through a dual-path residual mechanism (S-path and A-path). Specifically, The S-path filters out the refined information $\bm{f}_{j}\in \mathbb{R}^{T\times 2 \cdot F}$ from flow $\bm{p}$, formulated as $\bm{p}-\sum \bm{f}_{j}$, dynamically regulating the underlying information absorbed from flow $\bm{a}$. This gradually widening divergence between the two flows ensures the interaction of complementary information. Benefiting from the channel information integration capability of Harmonic Attention, we also employ it in each refinement $R$. In this way, $\bm{p}-\sum \bm{f}_{j}$ and $\bm{a}$ are concatenated in channel dimension, and fed into refinement $R_i$ to reconstruct the complete refined information. Through the A-path, the information flow $\bm{p}$ is skip-connected with each refined information $\bm{f}_{i}$ to form the final output of RUI $\bm{p}+\sum \bm{f}_{i}$. The inclusion of skip connection in the A-path ensures direct optimization of each flow. 
\vspace{-1em}  
\subsection{Auditory Constraint.}
The integration of the articulatory attributes into the framework, along with the incorporation of corresponding auditory perceptual attributes in the loss function, can be viewed as mutually reinforcing elements. Therefore, we introduce a combination of time-domain sample-level SI-SNR \cite{isik16_interspeech} and human psychoacoustic perception-based PMSQE \cite{martin2018deep} as the loss function $\mathcal{L}_{AC} = \mathcal{L}_{SI-SNR}+\mathcal{L}_{PMSQE}$ for the entire framework. This ensures the gradient-based optimization encompasses not only the numerical approximation but also the additional constraints of auditory masking and threshold effects as another kind of underlying information.
\section{Experiments}
\label{sec:experiment}
\subsection{Experimental Setup}
We conduct experiments on the 2020 DNS Challenge \cite{reddy20_interspeech}, which has 500 hours of clean speech from 2150 speakers and over 180 hours of noise from 150 classes. We generate 100 hours of noisy speech for the ablation and flexibility experiments, and 300 hours for the final comparison, respectively. The signal-to-noise ratio (SNR) ranges from -5 dB to 20 dB. The dataset is partitioned into a training set and a validation set in a 4:1 ratio. For testing audio, the SNR is between -5 dB and 30 dB, comprising a total of 10 hours of noisy speech. 

The 32 ms Hanning window with 25$\%$ overlap and 512-point STFT are used. The channel number of each refinement is 14. Model training is conducted using PyTorch with the Adam optimizer. The initial learning rate is set to 0.001, and a 0.75 learning rate decay will be applied if the validation loss does not decrease for 3 consecutive epochs. To provide a comprehensive assessment, we employ the number of the parameters (Para.), perceptual evaluation of speech quality (PESQ) \cite{rix2001perceptual}, scale-invariant signal-to-distortion ratio (SI-SDR) \cite{le2019sdr}, and short-time objective intelligibility measure (STOI) \cite{jensen2016algorithm}. For the last three metrics, higher values indicate better performance.

\subsection{Ablation Studies}
We utilize DPCRN \cite{DPCRN} as the PEM for ablation studies. As shown in TABLE~\ref{tab:1}, with the number of refinements $i$ increasing from 1 to 4, a trend of initially rising followed by a slight decrease is observed in the objective evaluations. This can be attributed to the S-path of multiple refinement iterator, where the remaining information obtained from flow $\bm{a}$ of Eq.~(\ref{7}) tends to become saturated. Concretely, the effective correction for the acoustic structure becomes limited. It is noteworthy that different PEMs lead to diverse information flows $\bm{p}$ of Eq.~(\ref{7}), consequently causing variations in the maximum value of $i$. 

Furthermore, we increase the number of convolutional channels in PEM to be comparable in parameters with RUI performing 4 times of refinements, denoted as PEM (large). The experimental result validates that blindly increasing the parameters of model is not advisable, the reasonable utilization of articulatory attributes is helpful to achieve better performance with fewer additional parameters. 

We also compare the auditory constraint (AC) with the SI-SNR loss function. The notable improvement underscores the complementarity of auditory and articulatory attributes in acoustic modeling. They reinforce each other within the established framework.

In addition, removing UIE results in an obvious drop in objective evaluations, which directly confirms the necessity of the introduced underlying information in our framework. Without the guidance of acoustic modeling based on articulatory attributes, the MRI cannot effectively perform structured correction on the output of PEM.
\begin{table}[htp]
	\centering
	\caption{Ablation studies.}
	\begin{tabular}{l|c|cccc}
		\toprule
&$i$ & Para. (M) & PESQ & SI-SDR (dB) & STOI ($\%$)  \\
\midrule
 \multirow{5}{*}{PEM (w/ $i$-ref)} & 0 & 1.64 & 2.771 & 20.389 & 94.40 \\
 &1 & 1.97 & 2.935 & 21.219 & 95.02 \\
 &2 & 2.25 & 2.941 & 21.262 & 95.14 \\
 &3 & 2.54 & 2.958 & 21.285 & 95.06 \\
 &4 & 2.82 & 2.909 & 21.113 & 94.92 \\
 \midrule
PEM (large) &0 & 2.95 & 2.774 & 20.380 & 94.39 \\
\midrule
RUI  &3 & 2.54 & 3.072 & 21.520 & 95.36 \\	
\quad - AC & 3 & 2.54 & 2.958 & 21.285 & 95.06 \\
 \quad \quad - UIE & 3 & 2.47 & 2.687 & 19.943 & 93.96 \\
\bottomrule
	\end{tabular}
	\label{tab:1}
\vspace{-2em} 
\end{table}
\subsection{Flexibility of RUI}
Besides DPCRN, we additionally select two spectrum estimation strategies for further exploration, including the baseline method NSNet \cite{xia2020weighted} from the 2020 DNS challenge and CRN \cite{tan2019complex}, a complex spectral mapping method. NSNet optimizes only the magnitude, while CRN estimates both the real and imaginary parts to enhance the magnitude and phase responses of noisy speech. As shown in TABLE~\ref{tab:2}, the flexibility of RUI enables it to better adapt to different PEMs. Specifically, it reports that the extent of improvement $\Delta$ exhibits variations. The retention of noisy phase in NSNet makes its supremum $I(\mathbb{P};\mathbb{C})$ lower than CRN. After refinement under the same condition, the lower supremum of NSNet, acting on the condition term $\mathbb{P}$ of underlying information $I(\mathbb{A};\mathbb{C}|\mathbb{P})$, leads to a higher extent of improvement $\Delta$. However, the obtained underlying information necessitates a greater trade-off in ameliorating the matter caused by retaining the noisy phase. In the case of CRN, the obtained underlying information is utilized more sufficiently to finely recover the acoustic structure of clean speech, culminating in a higher upgraded upper bound $\Gamma_{u}$ of RUI.
\begin{table}[htp]
\vspace{-1.0em}
	\centering
	\caption{Flexibility experiments.}
	\begin{tabular}{l|cccc}
		\toprule
  & Para. (M) & PESQ & SI-SDR (dB) & STOI ($\%$)  \\
\midrule
NSNet  & 2.79 & 2.274 & 17.023 & 91.06 \\
RUI [NSNet] & 3.97 & 2.829 & 20.460 & 94.56 \\
\quad $\Delta$ & 1.18 & 0.555 & 3.437 & 3.50 \\
\midrule
CRN & 1.70 & 2.732 & 20.295 & 94.31 \\
RUI [CRN] & 2.88 & 3.034 & 21.304 & 95.23 \\
\quad $\Delta$ & 1.18 & 0.302 & 1.009 & 0.92 \\
\bottomrule
	\end{tabular}
	\label{tab:2}
\vspace{-3em} 
\end{table}
\subsection{Visual Analysis}
\begin{figure}[htbp]
    \vspace{-0.5cm}
    \centering
    \subfigure[Noisy]
    {
    \label{fig:subfig1}\includegraphics[width=0.22\textwidth,height=0.07\textheight]{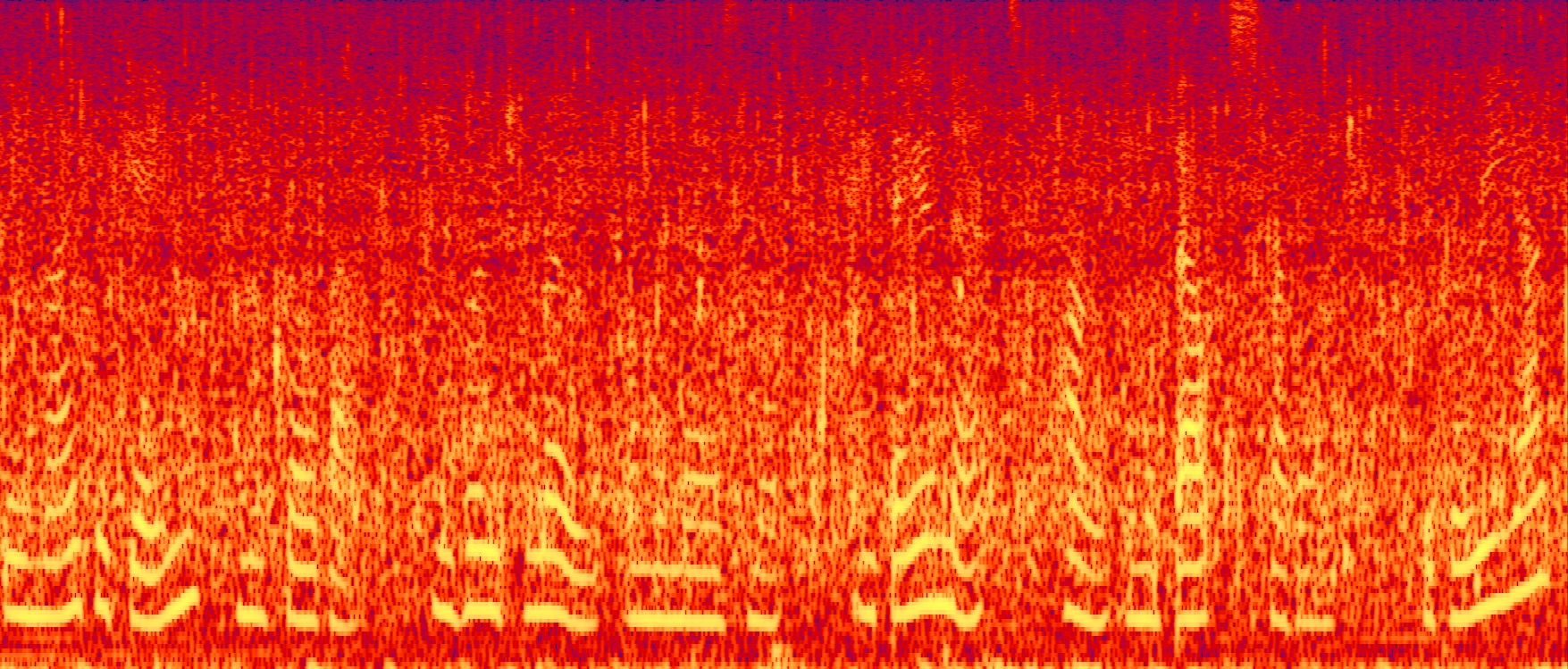}
      }
    \subfigure[Clean]
    {
      \label{fig:subfig2}\includegraphics[width=0.22\textwidth,height=0.07\textheight]{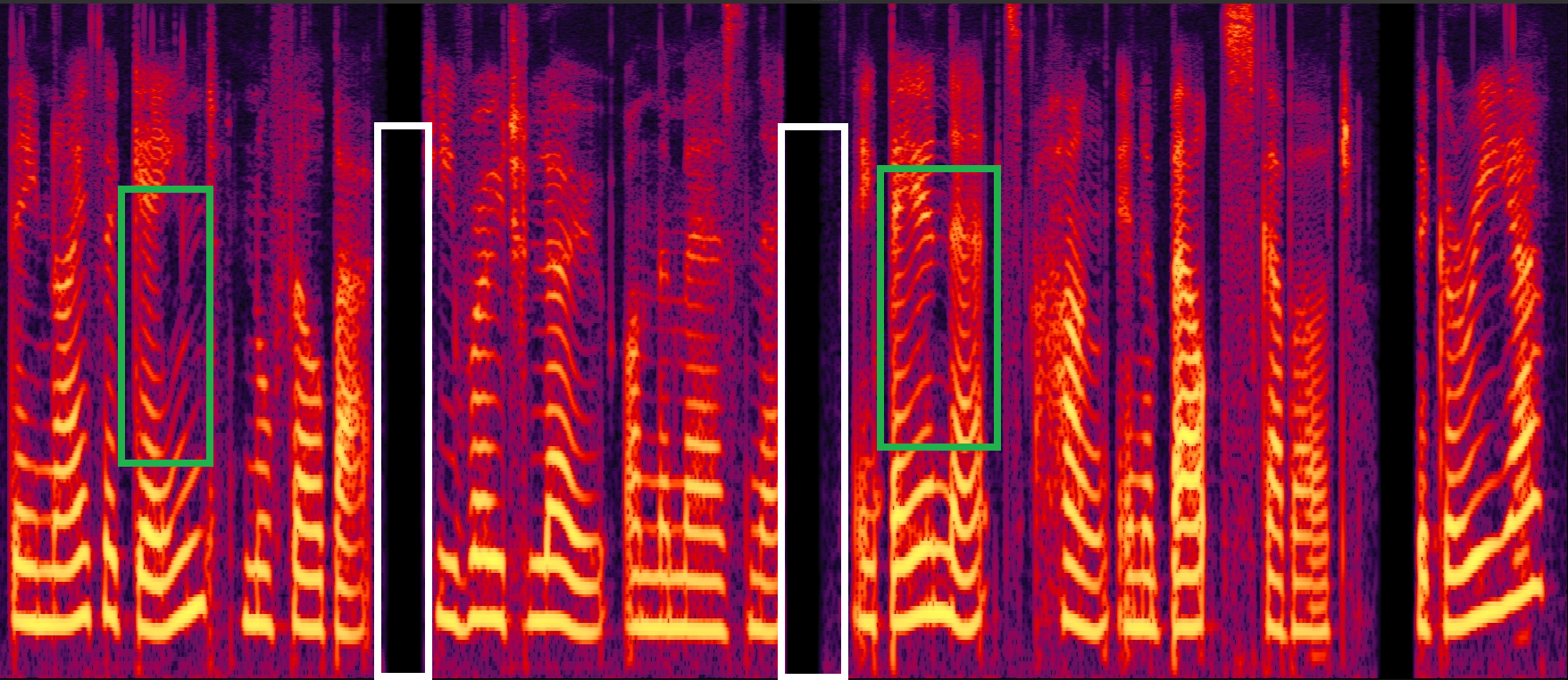}
      }
    \subfigure[PEM]
    {
      \label{fig:subfig3}\includegraphics[width=0.22\textwidth,height=0.07\textheight]{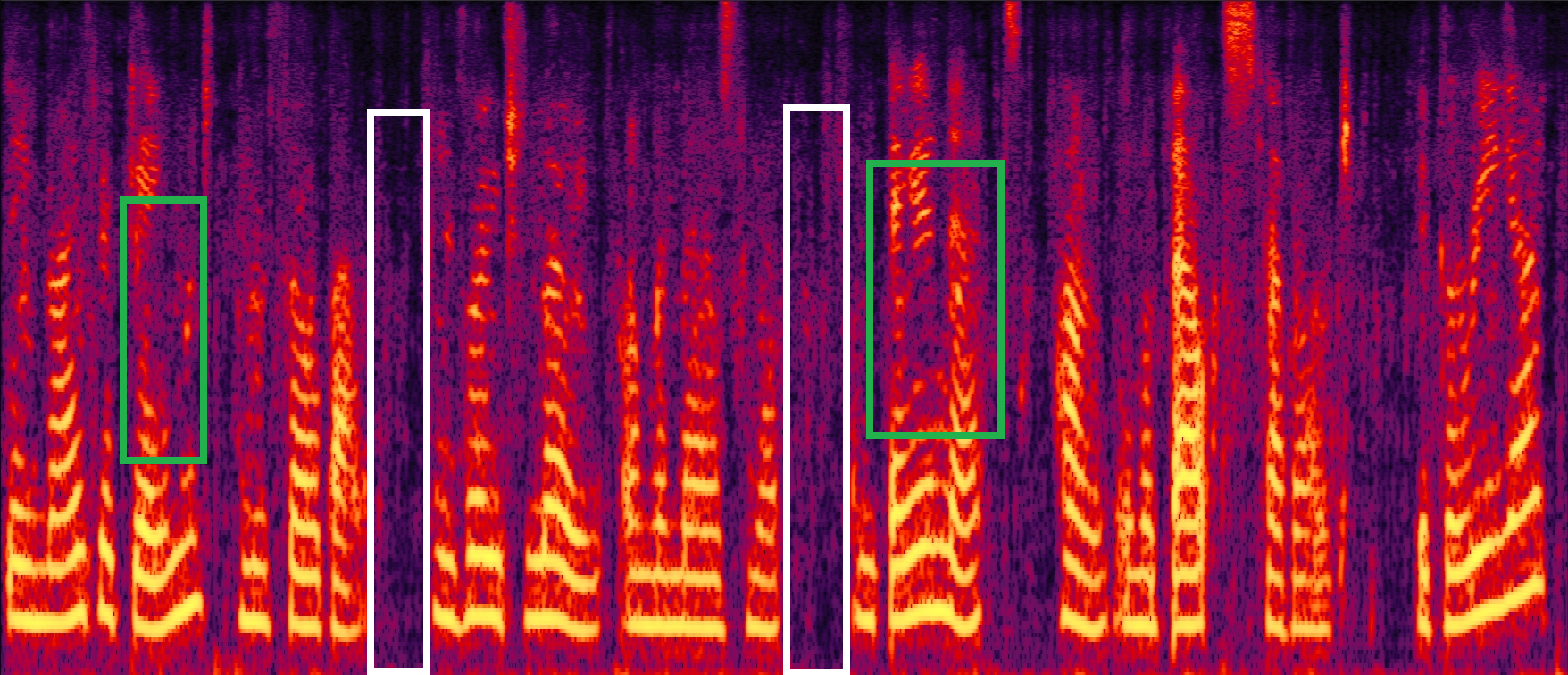}
      }
    \subfigure[RUI]
    {
      \label{fig:subfig4}\includegraphics[width=0.22\textwidth,height=0.07\textheight]{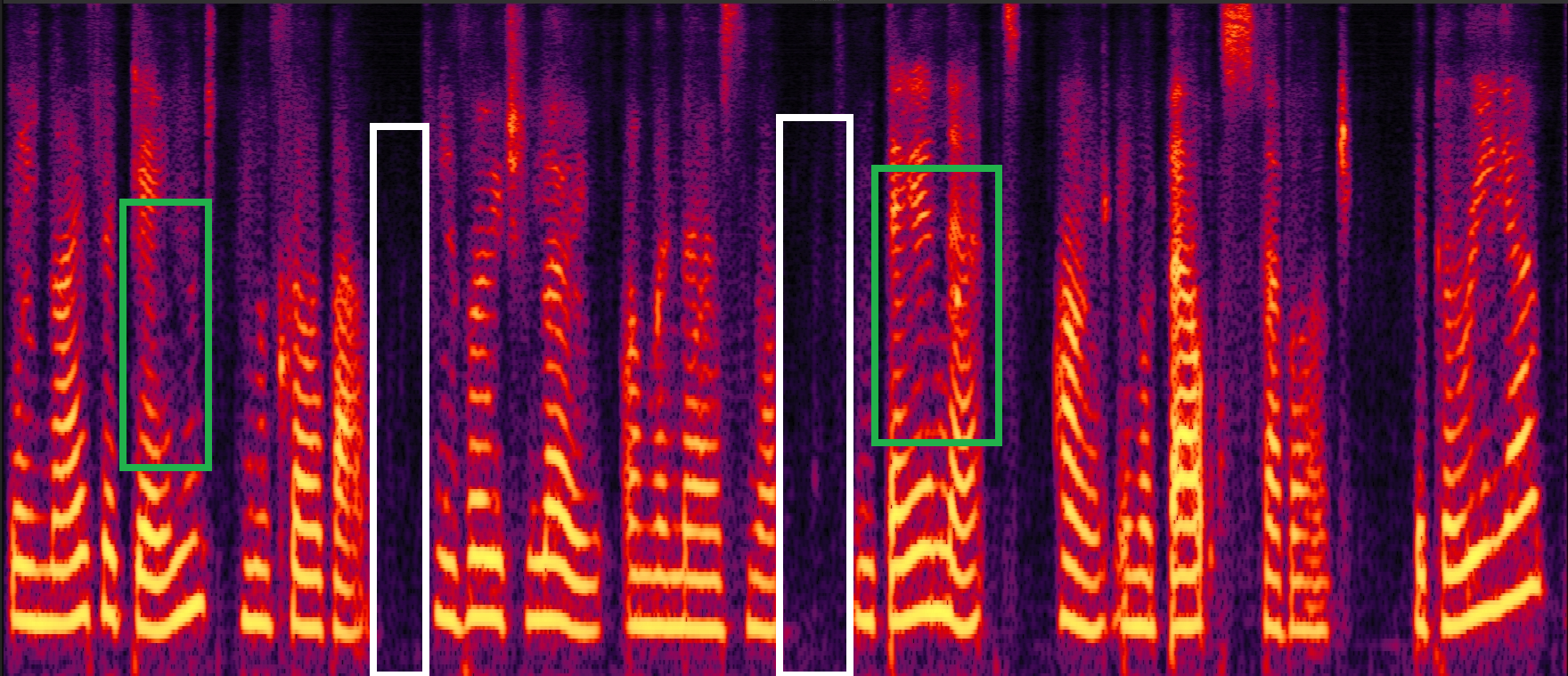}
      }
    \subfigure[1st-ref]
    {
      \label{fig:subfig5}\includegraphics[width=0.1\textwidth,height=0.05\textheight]{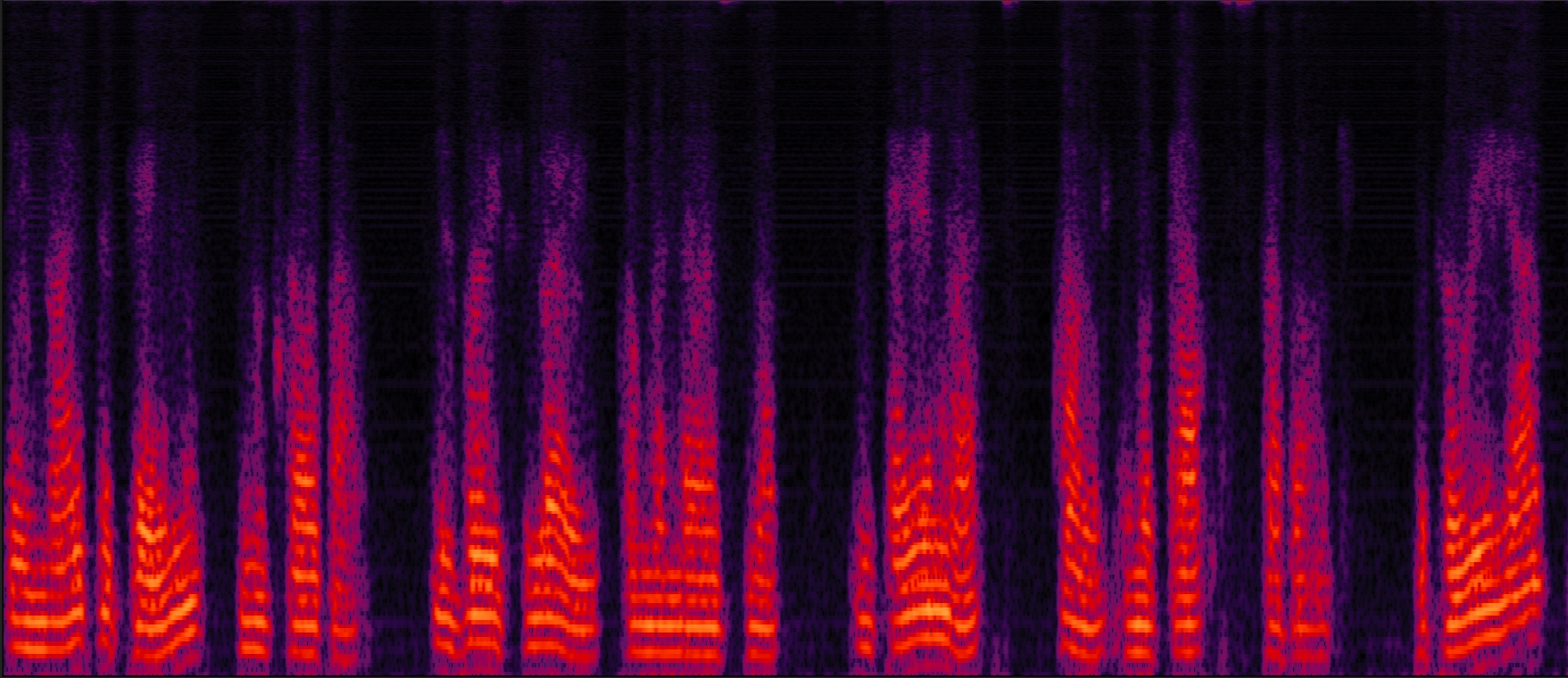}
      }
    \subfigure[2nd-ref]
    {
      \label{fig:subfig6}\includegraphics[width=0.1\textwidth,height=0.05\textheight]{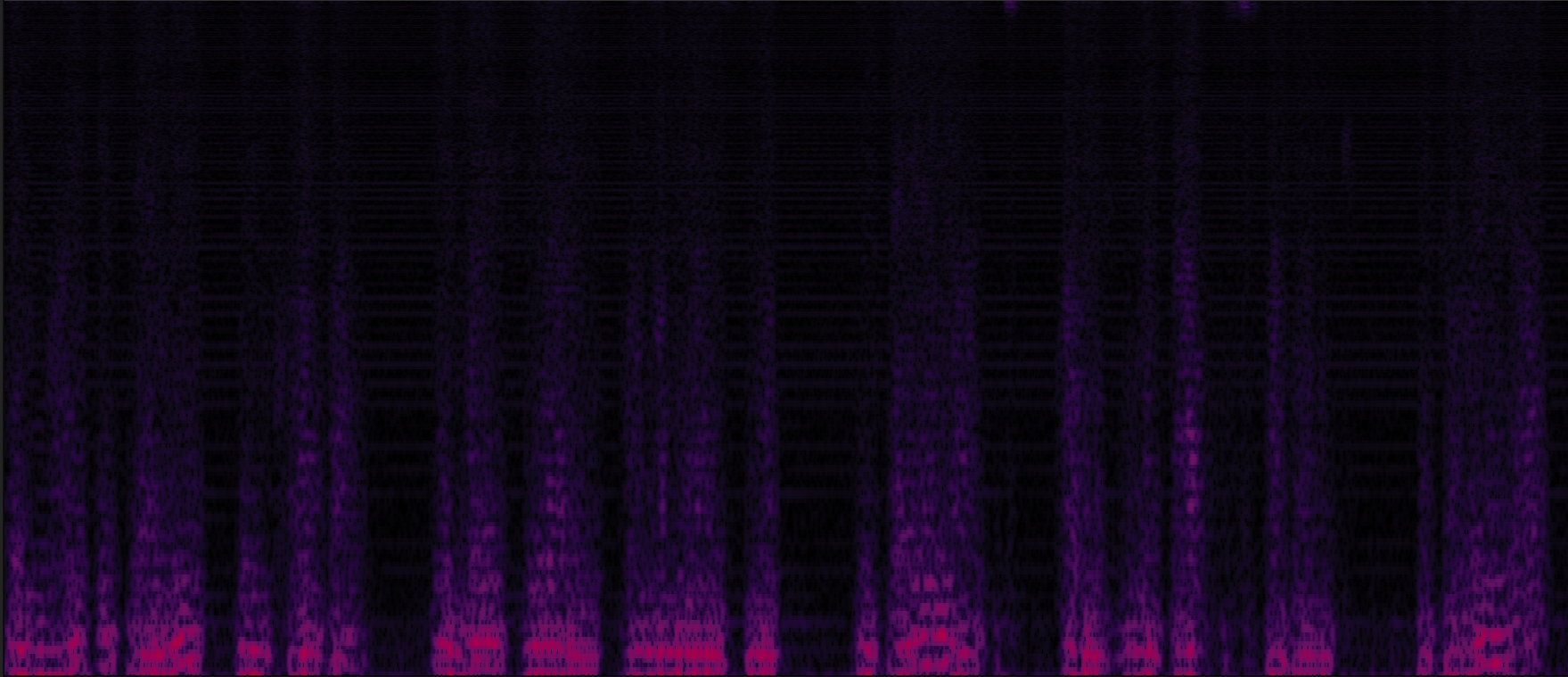}
      }
    \subfigure[3rd-ref]
    {
      \label{fig:subfig7}\includegraphics[width=0.1\textwidth,height=0.05\textheight]{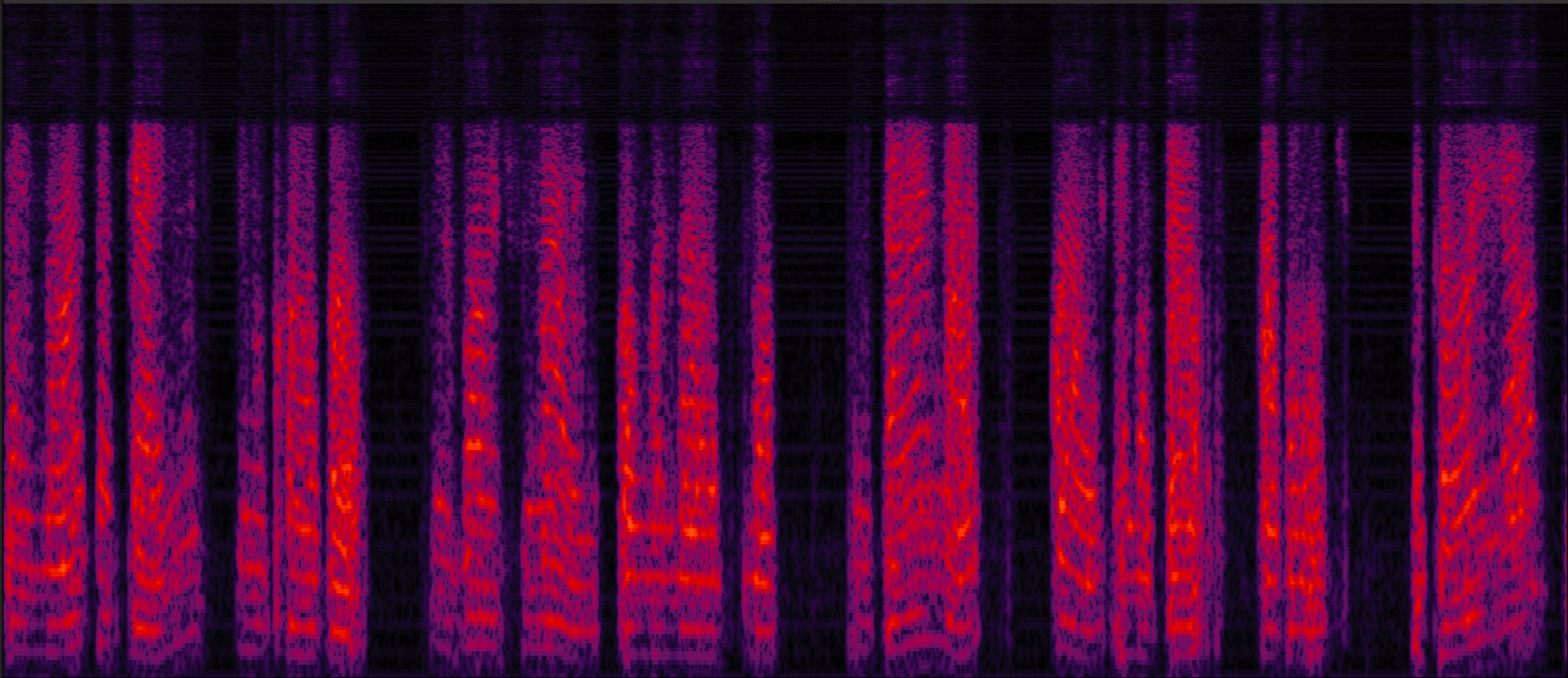}
      }
    \subfigure[4th-ref]
    {
      \label{fig:subfig8}\includegraphics[width=0.1\textwidth,height=0.05\textheight]{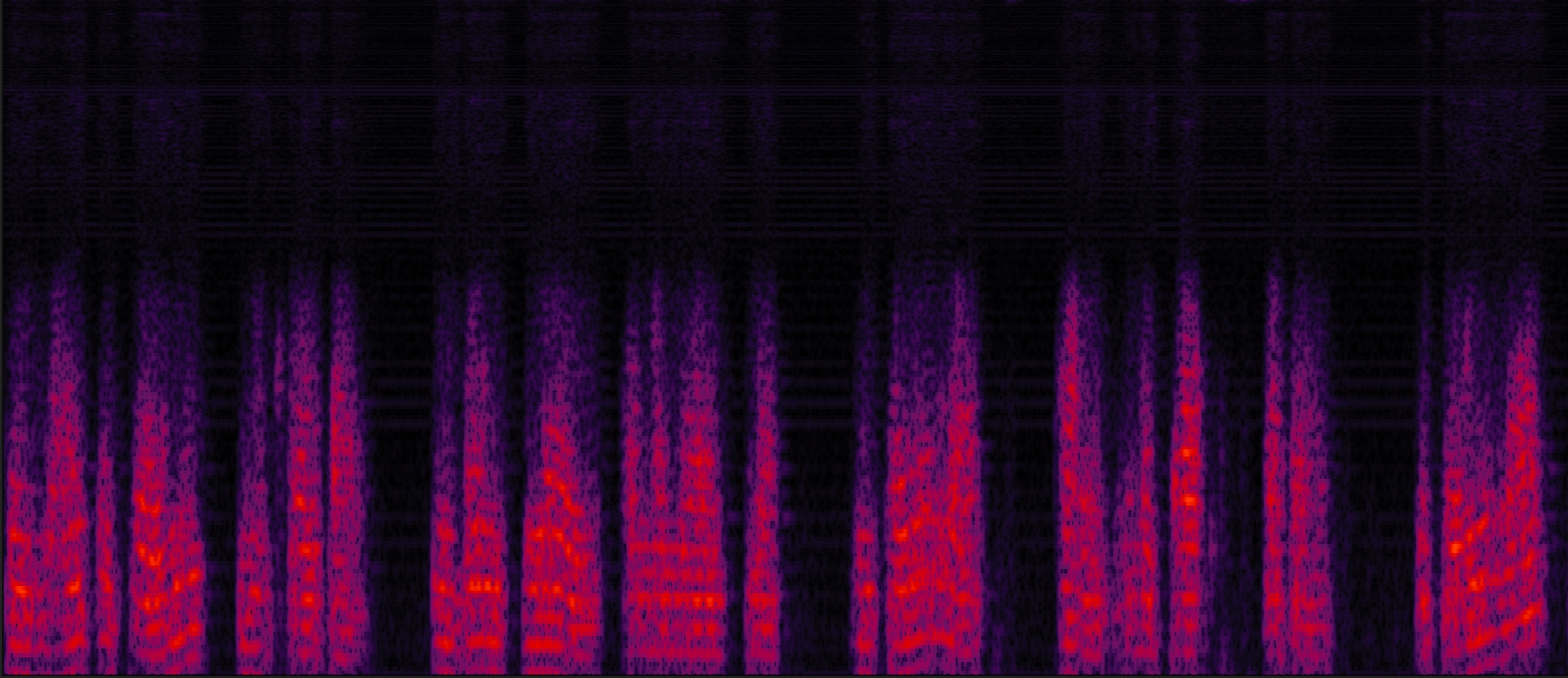}
      }
      \vspace{-0.8em}
      \caption{Noisy and clean spectrograms are provided as reference. (c)-(h) represent the output visualizations of each module and RUI, the $i$-th refinement is denoted as $i$th-ref.}
      \label{fig3}
\vspace{-2em}  
\end{figure}

To conduct a more in-depth analysis of information flow and the specific implication of refinement, we visualize the output of each module and RUI, in the final comparison, as shown in Fig.~\ref{fig3}. Despite the noticeable improvement in the output of PEM, there are still issues related to noise residual and speech degradation (colorful boxes in Fig.~\ref{fig:subfig3}). Fig.~\ref{fig:subfig4} shows that RUI not only removes the residual noise in the white box but also enhances the harmonic structure of speech in the green box, actively correcting the acoustic structural features of speech through articulatory attributes. The output of each refinement not only presents sparse representations of the spectrum, but also systematic harmonic structures, from the global level to different frequency bands.
\vspace{-1em}  
\subsection{Comparison on Public Test Dataset}
Compared to the outstanding solutions on the public test set of DNS-Challenge 2020, as shown in TABLE~\ref{tab:3}, our proposed RUI, using simple CRN as the backbone of PEM achieves competitive performance with minimal model parameters. The superiority of the proposed framework over the complex-spectrum methods (e.g., DCCRN/DCCRN+) is predictable. For the FullSubNet+ with 8.67 M parameters, RUI achieves better performance with only 33$\%$ of its parameters. For the HGCN, which only performs harmonic compensation, our solution shows comprehensive improvement in objective evaluations. It is also superior to the approach that combines DNN with a speech production model (GARNNHS). Moreover, in terms of the SI-SDR, our RUI outperforms an advanced method of hierarchical optimization in the complex spectrum (GAGNet), by a significant margin.
\begin{table}[htp]
\setlength\tabcolsep{1.5pt}
\vspace{-0.5em} 
	\centering
	\caption{System comparison on DNS-Challenge 2020 no reverb test set. “-” denotes no published result.}
	\begin{tabular}{l|ccccc}
		\toprule
 Model & Para. (M) & $\text{PESQ}_{\text{WB}}$ & $\text{PESQ}_{\text{NB}}$ & SI-SDR (dB) & STOI ($\%$)  \\
\midrule
Noisy & - & 1.58 & 2.45 & 9.07 & 91.52 \\
DCCRN \cite{hu20g_interspeech} & 3.70 & - & 3.27 & - & - \\
DCCRN+ \cite{lv21_interspeech} & 3.30 & - & 3.33 & - & - \\
FullSubNet \cite{hao2021fullsubnet} & 2.97 & 2.78 & 3.31 & 17.29 & 96.11 \\
FullSubNet+ \cite{chen2022fullsubnet+} & 8.67 & 2.98 & 3.50 & 18.34 & 96.69 \\
HGCN \cite{wang2022hgcn} & - & 2.88 & - & 18.14 & 96.50 \\
CARNNHS \cite{he2023speech} & - & 2.89 & 3.43 & 18.80 & 96.70 \\
GaGNet \cite{li2022glance} &5.94 & 3.17 & 3.56 & 18.91 & 97.13 \\
\midrule
RUI (Ours) & 2.88 & 3.02 & 3.50 & 19.54 & 97.11 \\	
\bottomrule
	\end{tabular}
	\label{tab:3}
\vspace{-2em} 
\end{table}
% \underline{http://graphicsqc.ieee.org/}
\section{Conclusion}
\label{sec:conclusion}
In this letter, we rethink the speech enhancement via Information Bottleneck principle, theoretically and practically. We point out that the prevalent noise suppression issues in existing methods stem from the incomplete restoration of the characteristics inherent in speech, especially acoustic characteristic. Theoretically, by defining the recovery of clean speech as incremental convergence of mutual information, we further express the acoustic characteristic of speech as conditional mutual information, (i.e., underlying information). Such a perspective can facilitate understanding and provide guidance for algorithmic design in speech enhancement. In practice, to ensure that the underlying information can be fully refined, we propose a universal framework called RUI with a dual-path residual mechanism, referring to the chain rule of entropy. Experimental results demonstrate that our solution has achieved highly competitive performance against other advanced methods, with a minimal number of parameters.
\bibliographystyle{IEEEtran}

\bibliography{mybib}

\end{document}